\documentclass[twocolumn]{aastex701}

\usepackage{amsmath,amssymb,bm}
\hypersetup{hidelinks}
\makeatletter
\renewcommand{\frontmatter@title@above}{}
\makeatother

\newcommand{\Athena}{\texttt{Athena++}}

\newcommand{\Mbh}{M_{\rm BH}}
\newcommand{\rg}{r_{\rm g}}

\newcommand{\etabg}{\eta_{\rm bg}}
\newcommand{\etaanom}{\eta_{\rm anom}}
\newcommand{\alphacrit}{\alpha_{\rm crit}}
\newcommand{\fcap}{f_{\rm cap}}
\newcommand{\qohm}{q_{\rm Ohm}}
\newcommand{\Qohm}{\dot{Q}_{\rm Ohm}}
\newcommand{\MdotBH}{\dot{M}_{\rm BH}}
\newcommand{\PhiBH}{\Phi_{\rm BH}}
\newcommand{\phiBH}{\phi_{\rm BH}}
\newcommand{\alphaJ}{\alpha_J}
\newcommand{\Sphi}{S_\Phi}

\shorttitle{Localized Current-Sheet Diffusion in M87-like GRMHD}
\shortauthors{Dai \& Cen}

\begin{document}

\title{A Localized Current-Sheet Magnetic-Diffusion and Heating Prescription\\
for Ideal-GRMHD Simulations of M87*-like
Accretion Flows}

\author{Songyan Dai}
\affiliation{Center for Cosmology and Computational Astrophysics, Institute for Advanced Study in Physics,
Zhejiang University, Hangzhou 310037, China}
\affiliation{Institute of Astronomy, School of Physics,
Zhejiang University, Hangzhou 310037, China}
\email[show]{soyonagasaki698@gmail.com}

\correspondingauthor{Renyue Cen}
\author{Renyue Cen}
\affiliation{Center for Cosmology and Computational Astrophysics, Institute for Advanced Study in Physics,
Zhejiang University, Hangzhou 310037, China}
\affiliation{Institute of Astronomy, School of Physics,
Zhejiang University, Hangzhou 310037, China}
\email[show]{renyuecen@zju.edu.cn}

\begin{abstract}
Ideal-GRMHD calculations provide the bulk accretion flows used in
black-hole emission models. A localized dissipation prescription
can add a resolved map of current-sheet activity while retaining
that dynamical reference. We implement such a prescription in
\Athena{}, using a smooth threshold on the dimensionless current
proxy $\alpha_J$ to select the magnetic-diffusion coefficient.
{The constrained-transport update is coupled to a conservative
energy-flux correction. Gas heating follows from primitive
recovery; $\qohm=\eta J^2$ is recorded as a diagnostic, without a
second energy source.} Static Fourier and traveling Alfv\'en tests
measure the evolved thermal gain, and an open-boundary
Schwarzschild test closes the Killing-energy ledger. The relative
total-energy residual is below $2\times10^{-15}$ in these tests.
A two-resolution Harris matrix gives localized rate exponents
$0.510\pm0.027$ and $0.488\pm0.028$, consistent with $1/2$ over
the specified early interval. The $\eta=10^{-3}$ rate exceeds the
same-resolution ideal control by a factor of 5.96.
In matched axisymmetric M87-like runs, localized diffusion changes
the early mean magnetic flux, accretion rate, and normalized flux
by less than $0.02\%$, with a radial-density distance of $0.005\%$.
{During the late active phase, a common-restart on/off comparison
keeps these mean shifts below $1.6\%$, within the intrinsic
temporal variability, and gives a density distance of $0.29\%$.}
Uniform diffusion at the tested coefficient reduces the raw flux
and accretion rate by $85\%$ and $99.8\%$. These comparisons
establish a spatially selective diffusion prescription with
explicit energy accounting and a quantified, small bulk-flow
response.
\end{abstract}

\keywords{accretion disks --- black hole physics ---
magnetic reconnection}

%-----------------------------------------------------------------------
\section{Introduction} \label{sec:intro}
%-----------------------------------------------------------------------

The Event Horizon Telescope (EHT) has resolved the millimeter shadow
of the supermassive black hole in M87 \citep{EHT2019PaperI}.
Interpreting the total-intensity and polarization images relies on
libraries of general-relativistic magnetohydrodynamic (GRMHD)
simulations \citep{EHT2019PaperV, EHT2021PaperVII,
EHT2021PaperVIII}; the underlying GRMHD codes have been compared on
common benchmarks \citep{Porth2019}. These simulations identify
current sheets and candidate reconnection sites
\citep{Ball2018, Ripperda2022}. A spatially resolved dissipation
field with explicit energy accounting connects those structures to
heating and radiative calculations while retaining the established
ideal-GRMHD bulk evolution.

Reconnection mediates horizon-flux evolution in MAD accretion flows
and plasma-filled black-hole magnetospheres
\citep{Ripperda2022, Bransgrove2021}. Radiative models infer
electron temperatures from local plasma conditions
\citep{Moscibrodzka2016}, and two-temperature simulations use
electron-heating prescriptions \citep{Chael2019}. GRAVITY detected
orbital motion during a near-infrared flare from Sgr~A*
\citep{GravityCollab2018}; MAD simulations connect comparable flare
phenomenology to intermittent reconnection \citep{Dexter2020}.
Resistive-GRMHD disk calculations likewise form current sheets,
plasmoids, and hot spots \citep{Ripperda2020}. These applications
motivate an explicit mapping from intense current structures to
local energy deposition.

Maxwell--Ohm solvers evolve the electric field through Amp\`ere's law
in special relativity \citep{Komissarov2007, Mignone2019}.
\citet{Palenzuela2009} developed an IMEX treatment of the
relativistic stiff relaxation. A general-relativistic implementation
is described by \citet[ApJS 244, 10]{2019ApJS..244...10R}. GRMHD accretion
codes also employ magnetic-diffusivity implementations
\citep{Qian2017}, and \citet{Bucciantini2013} developed a covariant
mean-field dynamo closure. Three-dimensional ideal and resistive
GRMHD simulations have examined current-sheet formation, plasmoid
evolution, and magnetic dissipation in multi-loop accretion flows
\citep{Nathanail2022}. Global uniform-resistivity calculations show
that sufficiently large resistivity alters magnetic-flux
accumulation and accretion-flow variability \citep{Nathanail2025}.
The spatial support of the non-ideal term is thus a physical design
choice for global accretion calculations.

\citet{Ball2018} characterized the plasma beta, magnetization, and
guide-field strength of candidate current sheets in ideal-GRMHD
accretion flows. At substantially higher resolution,
\citet{Ripperda2022} resolved plasmoid-mediated reconnection in a
three-dimensional ideal-GRMHD MAD at a rate of order
$0.01\,v_A$. Particle-in-cell simulations find rates of order $0.1\,v_A$
\citep{Sironi2014, Werner2018}. \citet{Moran2025} fitted a local
effective-resistivity prescription to fully kinetic simulations, and
\citet{Ripperda2026} applied that closure in resistive relativistic
MHD to recover comparable rates in local sheets and global
black-hole magnetospheres.

In phenomenological fluid models, unresolved kinetic effects
associated with fast reconnection are often represented through
anomalous resistivity. Threshold-type anomalous-resistivity models
and their kinetic motivation are reviewed by \citet{Yamada2010}.
Current-dependent nonuniform resistivity has also produced localized,
fast reconnection in relativistic merging-flux-tube calculations
\citep[MNRAS 485, 299]{2019MNRAS.485..299R}.
Together, these studies define a focused numerical problem:
construct a current-sheet dissipation diagnostic with controlled
local diffusion, explicit energy bookkeeping, and a measured impact
on the resolved bulk flow.

We solve this problem in \Athena{} \citep{Stone2020}. The
prescription uses $\alphaJ\propto |J|/B$ to select the high-current
tail, applies a metric-aware diffusion update, and records
$\eta J^2$ together with the conserved-energy flux correction.
The evaluation covers current-sheet locality, Harris-sheet
local-rate scaling and numerical-floor separation, thermal-energy
accounting, and preservation of the ideal-GRMHD bulk solution.
Matched ideal, uniform, and localized M87-like simulations
show that spatial localization supports intermittent
heating while preserving the global accretion state. Throughout,
$G=c=\Mbh=1$.

%-----------------------------------------------------------------------
\section{Numerical Method} \label{sec:methods}
%-----------------------------------------------------------------------

\subsection{Magnetic diffusion prescription in \Athena}
\label{sec:methods:ohmic}

The module augments the standard ideal-GRMHD state with a
coordinate-frame magnetic-diffusion operator. Relativistic
fluid-frame Ohm laws provide context for conductive dissipation
\citep{BlackmanField1993}; the prescription below is defined by its
spatial curl, CT edge increment, and energy-flux correction.

\begingroup 
We use metric signature $(-,+,+,+)$ and absorb $\sqrt{4\pi}$ into
the magnetic field. Let $g=\det(g_{\mu\nu})$,
$\gamma=\det(\gamma_{ij})$, and
$\alpha=(-g^{00})^{-1/2}$, so that
$\sqrt{-g}=\alpha\sqrt{\gamma}$. Throughout this paper,
$B^i\equiv{}^*F^{i0}$ denotes the native coordinate magnetic field
of \Athena{} \citep{2016ApJS..225...22W}. The Eulerian field is
$B_{(n)}^i=\alpha B^i$. Thus
\begin{equation}
 \sqrt{-g}\,B^i=\sqrt{\gamma}\,B_{(n)}^i
 \label{eq:magnetic_mapping}
\end{equation}
represents the same magnetic flux density in either convention.
The coordinate-field induction equation uses $\sqrt{-g}$,
consistently with the surface fluxes in Section~\ref{sec:m87}.

The stored fluid velocity is the spatial projection
$\tilde u^\mu=(\delta^\mu{}_{\nu}+n^\mu n_\nu)u^\nu$,
where $n_\mu=(-\alpha,0,0,0)$. With shift $\beta^i$, the conversion
to the transport velocity is
\begin{align}
 W&=\sqrt{1+\gamma_{ij}\tilde u^i\tilde u^j},
 &u^0&=W/\alpha,\nonumber\\
 u^i&=\tilde u^i-W\beta^i/\alpha,
 &v^i&=u^i/u^0.
 \label{eq:velocity_mapping}
\end{align}

The alternating symbol is $[123]=+1$ and the spatial Levi-Civita
tensor is $\epsilon^{ijk}=[ijk]/\sqrt{\gamma}$. The implemented
current proxy and its norms are
\begin{align}
 B_i&=\gamma_{ij}B^j, &B^2&=\gamma_{ij}B^iB^j,\nonumber\\
 J^i&=\frac{[ijk]}{\sqrt{\gamma}}\partial_jB_k,
 &J^2&=\gamma_{ij}J^iJ^j.
 \label{eq:current}
\end{align}
The curl acts on the native $B^i$.
Table~\ref{tab:variable_mapping} specifies the storage locations.
\endgroup 

The GR path first interpolates the face field to cell centers,
lowers its index, and takes centered coordinate derivatives.
It then averages $J^i$ and the spatial metric to the adjacent
edges before contracting to obtain $J_i$. The edge coefficient
$\eta_e$ is the arithmetic mean of the adjacent cell coefficients.
The added edge-array value is $e_{\eta,e}=\eta_e(J_i)_e$ for an
edge in direction $i$. \begingroup The cell diagnostic is
\begin{equation}
 \qohm=\eta_c\,\gamma_{ij,c}J_c^iJ_c^j,
 \label{eq:qohm}
\end{equation}
computed from the cell-centered current before edge interpolation.
\endgroup 

\begin{table*}[t]
\caption{{Conventions and native array mapping.}\label{tab:variable_mapping}}
\centering
\begin{tabular}{lll}
\hline\hline
Quantity & Definition & Storage\\
\hline
$B^i$ & ${}^*F^{i0}$ & face-area average, \texttt{b.x1f/x2f/x3f} \\
$B_c^i$ & face-to-center interpolation & \texttt{bcc(IB1/IB2/IB3)} \\
$\rho,p_{\rm gas}$ & fluid-frame density and pressure & \texttt{w(IDN/IPR)} \\
$\tilde u^i$ & normal-frame spatial projection & \texttt{w(IVX/IVY/IVZ)} \\
$J_c^i$ & Equation~(\ref{eq:current}) & \texttt{fdif.jcc(IB1/IB2/IB3)} \\
$e_{\eta,e}$ & $\eta_e(J_i)_e$ & edge-centered \texttt{e\_oa.x1e/x2e/x3e} \\
$U_E$ & $T^0{}_0$ in the GR build & cell-volume average, \texttt{u(IEN)} \\
$\qohm$ & Equation~(\ref{eq:qohm}), diagnostic & cell-centered user output \\
\hline
\end{tabular}
\parbox{0.94\textwidth}{The coordinate volume is
$V_c=\int_c\sqrt{-g}\,d^3x$. A proper spatial volume uses
$\sqrt{\gamma}$ instead. Neither magnetic array is densitized.}
\end{table*}

\subsection{Discrete induction and energy bookkeeping}
\label{sec:methods:energy}
\begingroup 
For one stage of duration $\Delta t_s$, let a star denote the
stage input including the ideal contributions. The diffusion
increment applied to a face is
\begin{equation}
 B_f^{n+1}=B_f^*-
 \frac{\Delta t_s}{A_f}
 \sum_{e\in\partial f}s_{fe}L_e e_{\eta,e},
 \label{eq:ct_discrete}
\end{equation}
where $s_{fe}$ fixes the oriented circulation. Here $A_f$ and
$L_e$ are the face and edge integration weights returned by the
coordinate module. In the GR implementation they integrate
$\sqrt{-g}$ over the corresponding coordinate face or edge;
$L_e$ is not a proper spatial length. In the differential limit,
\begin{align}
 \partial_t(\sqrt{-g}B^i)
 &=-\partial_j\!\left[\sqrt{-g}(v^jB^i-v^iB^j)\right]\nonumber\\
 &\quad-\partial_j\!\left([ijk]\sqrt{-g}\,e_{\eta,k}\right).
 \label{eq:induction_native}
\end{align}

The face correction computed by the diffusion module is
\begin{equation}
 P_{\eta,f}^i=\frac{[ijk]}{\sqrt{\gamma_f}}
             \overline e_{\eta,j,f}\,\overline B_{k,f}.
 \label{eq:pflux_code}
\end{equation}
The transverse EMFs are averaged from edges to the face; the
magnetic field is averaged from the two neighboring cell centers
and lowered with the face metric. We add
$F_{E,\eta,f}^i=-P_{\eta,f}^i$ to the native \texttt{IEN} flux. The discrete
energy increment is
\begin{equation}
 U_{E,c}^{n+1}=U_{E,c}^{*}
 -\frac{\Delta t_s}{V_c}
       \sum_{f\in\partial c}s_{cf}A_fF_{E,\eta,f}
 +\Delta U_{E,c}^{\rm floor},
 \label{eq:energy_discrete}
\end{equation}
with no independent $\Delta t_s\qohm$ source. Gas internal energy
is obtained through primitive recovery from the updated conserved
variables and magnetic field. Floor or recovery repairs belong to
the separate last term.

The physical energy associated with stationary time translation is
$E_K=-\sum_c V_cU_{E,c}$. In Minkowski coordinates,
$U_E=-T^{00}$, so the outward Poynting flux $S^i=P_\eta^i$
corresponds to the stored energy flux $-S^i$. The Killing-energy
ledger uses the face flux in Equation~(\ref{eq:pflux_code}).
The coordinate diagnostic
$\eta J^2$ supplies a dissipation weight, with a normalization
distinct from a fluid-frame heating scalar.
\endgroup 

\subsection{Localized current-sheet trigger model}
\label{sec:methods:anom}

The total Ohmic resistivity at each cell is
\begin{equation}
\eta(\mathbf{x}, t) =
\etabg
+ \etaanom
  \frac{1}{2}\left[1 + \tanh\!\left(
  \frac{\alphaJ-\alphacrit}{\delta\alpha}\right)\right],
\label{eq:trigger}
\end{equation}
where
\begin{equation}
\alphaJ \equiv \ell_0\frac{|J|}{B}
\label{eq:alpha_trigger}
\end{equation}
is dimensionless. We use $\ell_0=\rg$ for the M87 runs and the
normalized problem length for the flat-space tests; both equal one
in their respective code units. The parameter $\alphacrit$ is the
trigger threshold, and $\delta\alpha$ controls the transition width.
In the bulk flow ($\alphaJ\ll\alphacrit$), the anomalous
coefficient decreases exponentially toward the background value,
$\etabg=0$ in the localized runs. The ideal G0 calculation provides
the common numerical-dissipation reference. In intense current sheets
($\alphaJ \gg \alphacrit$)
the resistivity saturates at $\etabg+\etaanom$.

{The code evaluates
$B=\sqrt{\max(\gamma_{ij}B^iB^j,0)}$ and
$|J|=\sqrt{\max(\gamma_{ij}J^iJ^j,0)}$, and replaces the
denominator in Equation~(\ref{eq:alpha_trigger}) by
$\max(B,10^{-20})$ in code units. Cells with $B=|J|=0$
have $\alphaJ=0$ and $\qohm=0$.}

Motivated by threshold-type models reviewed by
\citet{Yamada2010}, we select intense current concentrations with
a fixed empirical threshold. The M87 comparison uses
$\alphacrit=85.1$ and $\delta\alpha=8.5$ throughout the early
and late experiments. The scan $\alphacrit\in\{70,85.1,100,115\}$
measures sensitivity to that choice
(Section~\ref{sec:results:robust}).

A per-step cap bounds the diagnostic increment $\qohm\Delta t$
by a fraction $\fcap$ of the fluid-frame internal-energy density:
\begin{equation}
\eta \leq \eta_{\rm cap}
\equiv \frac{\fcap e_{\rm int}}{J^2\Delta t}.
\label{eq:qcap}
\end{equation}
The runs use $\fcap=0.2$, with $0.1$ and $0.4$ included in the
robustness scan.

\subsection{Verification and validation suite}
\label{sec:vandv}

The tests measure magnetic diffusion, gas-energy gain, and
total-energy balance, recording any primitive-recovery corrections.

\textbf{A1.} Periodic Fourier diffusion starts with
$B^y=B_0\sin(2\pi x)$, $B_0=0.05$, $\rho=p=1$, and zero velocity
on $0\leq x\leq1$. We use $\eta=10^{-2}$, $\Gamma=5/3$, and
$t_{\rm end}=0.2$. {At $N_x=256$, the magnetic-energy loss is
$9.139\times10^{-5}$ and the recovered internal-energy gain is
$9.132\times10^{-5}$. The latter agrees with the cumulative
$\eta J^2$ to $0.054\%$. A separate single-step calculation at
$\Delta t=10^{-5}$ checks the spatial distribution of the heat:
$\Delta e_{\rm int}/\Delta t$ agrees with
$\eta(2\pi B_0)^2\cos^2(2\pi x)$ to $1.76\times10^{-4}$ in
relative $L_1$ norm.}

\textbf{A2.} A traveling Alfv\'en perturbation tests energy
transfer in a moving relativistic fluid. The wave uses $B^x=0.5$,
$\rho=p=1$, $\Gamma=5/3$, and transverse velocity amplitude $0.01$,
with the enthalpy-dependent polarization given in the Appendix.
{At $t=2$, the evolved internal-energy gain is
$1.3591\times10^{-4}$. Removing the reversible contribution using
the initial uniform entropy gives an irreversible thermal excess
of $1.02344\times10^{-4}$, compared with
$I_Q=1.02336\times10^{-4}$: a relative difference of
$7.98\times10^{-5}$. The same wave with $\eta=0$ gives an
irreversible excess of $1.30\times10^{-8}$.}

\textbf{A3.} An error-function current sheet $B^y(x,0) = B_0\,
\mathrm{erf}((x-x_0)/a_0)$ broadens self-similarly as
$a(t) = {\sqrt{a_0^2 + 4\eta t}}$.

\begin{figure*}[t]
\centering
\includegraphics[width=0.98\textwidth]{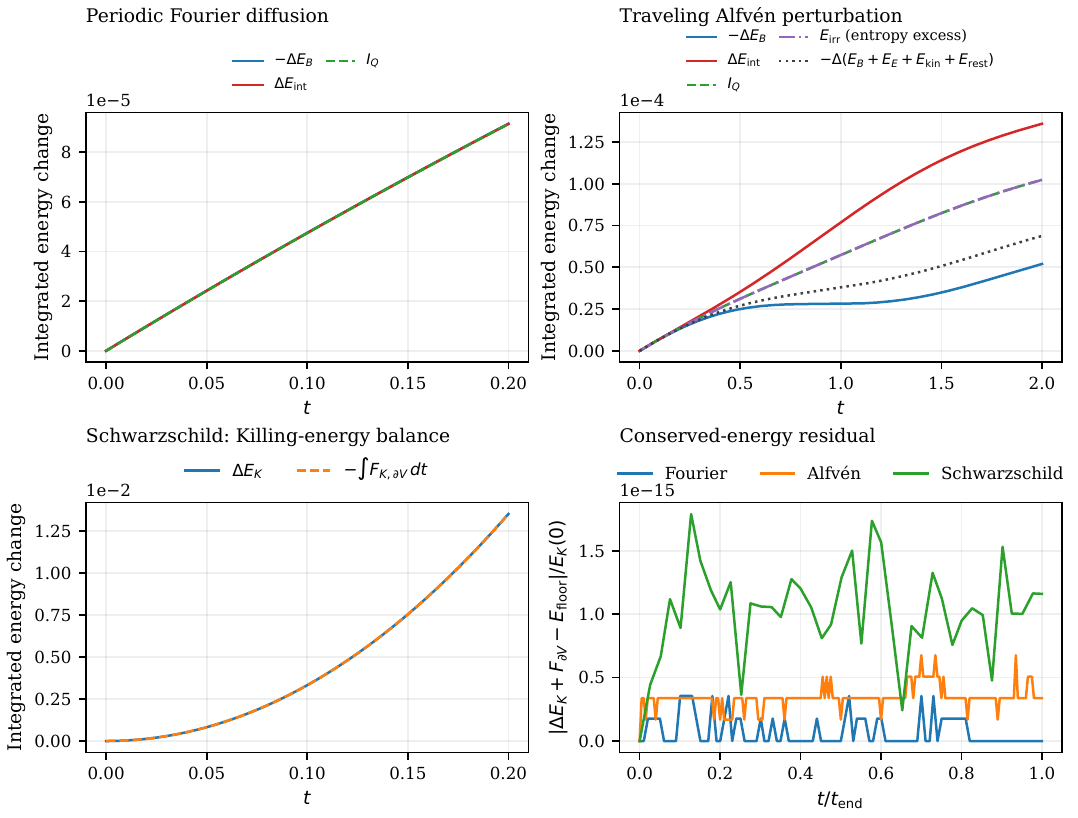}
\caption{{Evolved energy-transfer ledger.} Top left: Fourier
magnetic loss, recovered internal-energy gain, and cumulative
$\eta J^2$ at $N_x=256$. Top right: the traveling Alfv\'en test,
including the sum of mechanical and electromagnetic energy changes
and the entropy-based irreversible thermal excess. Bottom left:
the Schwarzschild Killing-energy change and the negative integrated
boundary energy flux. Bottom right: the relative conserved-energy
residual for all three tests. No floor repair occurs in these runs.
The Alfv\'en internal-energy increase includes reversible
compression; its entropy excess follows $I_Q$.}
\label{fig:energy_budget}
\end{figure*}

\textbf{A4.} Figure~\ref{fig:energy_budget} presents the cumulative
energy ledger for A1, A2, and the Schwarzschild test A5. For the
periodic tests, {the boundary contribution cancels to roundoff.
No conserved-variable repair is applied at any predictor or
corrector stage. The largest relative total-energy residual is
$6.8\times10^{-16}$ for the two flat-space transfer tests.}
The ledger tests set $\rho_{\min}=10^{-10}$,
$p_{\min}=10^{-12}$, and $W_{\max}=50$, without
magnetization-dependent floors.

\begingroup 
For a static periodic Minkowski test, the cumulative ledger is
\begin{align}
 E_{\rm int}&=\sum_c V_c\frac{p_c}{\Gamma-1},&
 E_{\rm mag}&=\sum_c V_c\frac{B_c^2}{2},\nonumber\\
 I_Q(t)&=\int_0^t\!\sum_c V_c\eta_c J_c^2\,dt',&
 E_{\rm tot}&=-\sum_c V_c U_{E,c}.
 \label{eq:flat_ledger}
\end{align}
Here $p_c$ is recovered from the evolved state. The global residual is
\begin{equation}
 \mathcal R_E=\Delta E_{\rm tot}
 +\int_0^t\!\sum_{\partial V}F_{E,\rm phys}A\,dt'
 -\Delta E_{\rm floor}.
 \label{eq:ledger_residual}
\end{equation}
The face-flux integral and $I_Q$ use the VL2 corrector-stage quadrature.
The small kinetic response of A1 accounts for the difference
between magnetic loss and thermal gain. For A2, the relativistic
energy decomposition and the entropy excess separate irreversible
heating from compression (Appendix~\ref{sec:wave_ledger}).
\endgroup 

\begingroup 
\textbf{A5.} A discrete vector-potential curl initializes a
three-dimensional field in Schwarzschild coordinates. The initial
and maximum evolved CT divergence are $5.98\times10^{-18}$ and
$1.35\times10^{-16}$. The Killing-energy balance, including the
open-boundary flux, closes to $1.79\times10^{-15}$ relative to
the initial energy. Appendix~\ref{sec:a5} specifies the field,
grid, and constraint normalization.
\endgroup 

\begin{figure*}[!htbp]
\centering
\includegraphics[width=0.98\textwidth]{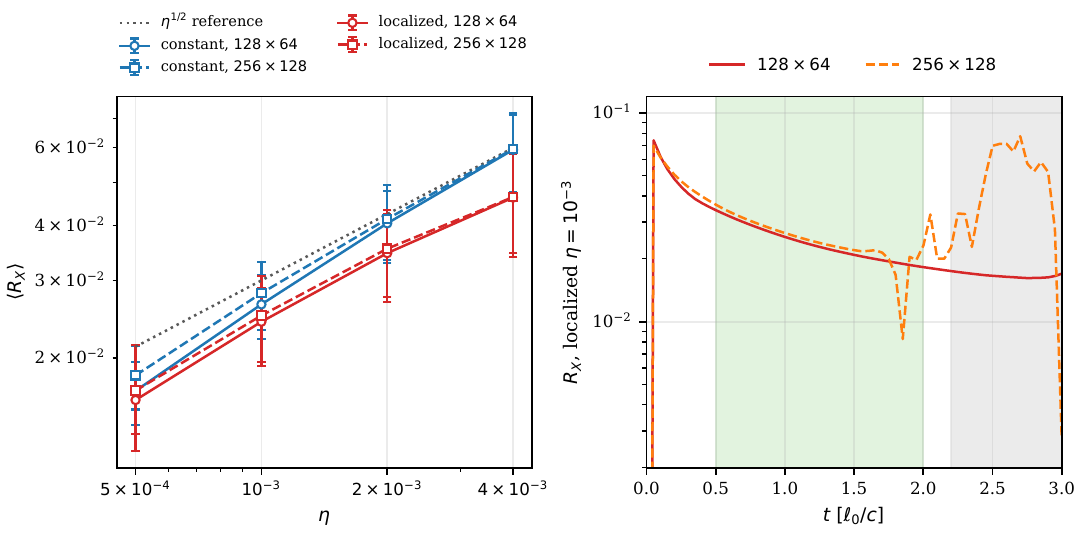}
\caption{Two-resolution Harris-sheet benchmark.  \emph{Left:}
local X-point reconnection rate, normalized by the fixed initial
far-upstream $B_{{\rm up},0}v_{A,0}$ and averaged over
$t=0.5$--$2.0$, for constant and localized resistivity.  Error
{bars show temporal standard deviations, not independent-sample
errors on the mean; quoted exponent uncertainties are the
one-standard-error scatter estimates from fitting four case means
in log--log space.} The dotted line is the
$\eta^{1/2}$ reference.  \emph{Right:} rate histories for the
localized $\eta=10^{-3}$ cases.  The green region marks the common
fit window. The late high-resolution excursions in the gray
$t=2.2$--$3.0$ interval are excluded from the primary fit.
Times are in code units $\ell_0/c$, and velocities and the
initial Alfv\'en normalization use the relativistic definitions
given in the text. The localized model uses Equation~(\ref{eq:trigger})
directly.}
\label{fig:sweet_parker}
\end{figure*}

We additionally run constant- and localized-resistivity
Harris-sheet benchmarks in Minkowski geometry at
$128\times64$ and $256\times128$, each with
$\eta \in \{5\!\times\!10^{-4},\,10^{-3},\,2\!\times\!10^{-3},\,
4\!\times\!10^{-3}\}$ (Figure~\ref{fig:sweet_parker}).  We measure
the standard local rate at the symmetry-fixed primary X-point,
\begin{equation}
 R_X = \frac{|E_z(X)|}{B_{{\rm up},0}v_{A,0}},
\end{equation}
{where $E_z(X)$ is a cell-reconstructed induction diagnostic,
$-v^xB^y+v^yB^x+\eta J_z$, sampled in the cell nearest the
symmetry-fixed primary X-point. It is not an output of the
CT edge EMF. We convert the stored projected velocities using
$v^i=\tilde u^i/\sqrt{1+\sum_j(\tilde u^j)^2}$.
The runs use the GRMHD module in Minkowski coordinates, with
$p=(\Gamma-1)e_{\rm int}$ and $\Gamma=5/3$. The initial fluid
is at rest, so the upstream normalization is
$v_{A,0}^2=B_{{\rm up},0}^2/(\rho_0h_0+B_{{\rm up},0}^2)$,
where $h_0=1+\Gamma p_0/[(\Gamma-1)\rho_0]$.
For the asymptotic background, $v_{A,0}=1/\sqrt{3.25}=0.555$;
the numerical sample uses its local initial values.} The fixed
normalization is measured from the initial state
at $|y|=10a=0.2$, sufficiently far upstream of the sheet.  Using
the common early analysis window
$t\in[0.5,2.0]$, \begingroup the localized prescription gives
\begin{equation}
 R_X\propto
 \begin{cases}
 \eta^{0.510\pm0.027}, & 128\times64,\\
 \eta^{0.488\pm0.028}, & 256\times128,
 \end{cases}
\end{equation}
consistent with the classical Sweet--Parker exponent $1/2$
within the fitted uncertainties.  The constant-$\eta$ controls
give $0.607\pm0.015$ and $0.568\pm0.015$, respectively, moving closer to the $1/2$ exponent with increasing resolution. \endgroup  At the
primary X-point the localized trigger is saturated, so the
measured local resistivity equals the nominal $\eta$ within
saved-output precision. Late high-resolution excursions change the
longer-window fit. We retain the earlier common interval and
quantify the sensitivity to its endpoints in the Appendix.
{A $1/2$ exponent also follows from an initially thin diffusing
sheet: $\delta\propto\sqrt{\eta t}$ gives
$\eta J\propto\sqrt{\eta}$ at fixed time. Our result establishes
local-rate exponent consistency, without uniquely identifying a
quasi-steady Sweet--Parker layer. Establishing that regime requires
layer geometry and relativistic mass, induction, and outflow
balances \citep{2005MNRAS.358..113L}.}

\begingroup 
The explicit signal is also separated from the numerical floor:
at $\eta=10^{-3}$, the mean rate over $t=0.5$--$2.0$ is 5.96
times that of the $\eta=0$ run on the same $128\times64$ grid (see
Appendix~\ref{sec:appendix_vnv}). Across all 16 finite-$\eta$ scan
runs, the magnetic constraint remains at
$\max|\nabla\!\cdot\!B|=4.09\times10^{-12}$.
\endgroup 

%-----------------------------------------------------------------------
\section{M87 Setup} \label{sec:m87}
%-----------------------------------------------------------------------

\subsection{Initial conditions and grid}

We initialize a Fishbone--Moncrief torus
\citep{FishboneMoncrief1976} around a Kerr black hole of mass
$M = 1$ and dimensionless spin $a_* = 0.5$ in horizon-penetrating
Kerr--Schild coordinates \citep{McKinney2004}, giving an event
horizon at $r_{\rm h} = 1 + \sqrt{1-a_*^2} \simeq 1.866\,\rg$. The
torus pressure maximum is at $r_{\rm max} = 12\,\rg$.
{The inner torus edge is $6\,\rg$, and the peak density is
normalized to one. A single-loop poloidal field is seeded with
$A_\phi=0.4178\max(\rho-0.2,0)$ in code units; this amplitude
sets the magnetic
normalization.}

The 2D axisymmetric grid has
$N_{r}\times N_{\theta} = 128\times 128$ in
$r \in [1.78,\,40]\,\rg$ {with uniform radial and polar
spacing and an inner boundary inside the horizon.}
The parameter-scan runs are evolved to
$t=500\,\rg/c$ with snapshot cadence $5\,\rg/c$. Activity statistics
use $80\leq t/(\rg/c)\leq500$, and the early bulk-flow comparison
uses $200\leq t/(\rg/c)\leq500$. An additional G3 fiducial run is extended to
$t = 1500\,\rg/c$ to characterize the long-term recurrent heating
phenomenology (Section~\ref{sec:results:longrun}). To measure the
late-time bulk response during this active phase, we evolve the
fiducial G3 model to $t=1000\,\rg/c$ and branch matched
continuations with the localized term on and off. Both
branches are evolved to $t=1500\,\rg/c$ and compared over
$t=1050$--$1500\,\rg/c$.

\begingroup 
Table~\ref{tab:numerical_setup} collects the numerical settings.
Radial user boundaries copy density, pressure, and transverse
projected velocity from the adjacent active zone. They clip the
radial projected velocity to $\tilde u^r\leq0$ at the inner boundary
and $\tilde u^r\geq0$ at the outer boundary. Transverse magnetic
components are extended with metric factors, and the normal face
field is extended using the discrete divergence constraint.
The principal diffusion tensor of the edge-slot convention in
Equation~(\ref{eq:induction_native}) is
$\eta\sqrt\gamma\,\gamma^{ij}$. We bound its explicit step using
\begin{align}
 \Lambda_c &= \sum_i\frac{\gamma^{ii}_c}{(\Delta x_c^i)^2}
       +2\sum_{i<j}\frac{|\gamma^{ij}_c|}{\Delta x_c^i\Delta x_c^j},\nonumber\\
 \Delta t_{\eta}&\leq C_{\rm CFL}\min_c
 \frac{1}{2\eta_{{\rm bound},c}\sqrt{\gamma_c}\,\Lambda_c},
 \label{eq:diffusion_timestep}
\end{align}
where the sums include the active spatial directions. The
inverse spatial metric is
$\gamma^{ij}=g^{ij}+\alpha^2g^{0i}g^{0j}$.
For the localized Ohmic term the bound uses
$\eta_{\rm bg}+\eta_{\rm anom}$, before the energy cap. The
step is the smaller of the hyperbolic and diffusion restrictions.
This reduces to the Cartesian diffusion bound on a uniform
orthogonal grid. The common M87 timestep is approximately
$7.86\times10^{-3}\,\rg/c$. The ideal and uniform controls reduce
their hyperbolic CFL factor to $0.06022034$ to match the localized
run, whose $C_{\rm CFL}=0.3$ diffusion bound sets this step.
\endgroup 

\begin{table*}[t]
\caption{{Numerical setup of the fiducial production runs
and Harris scaling matrix.}\label{tab:numerical_setup}}
\centering
\newcommand{\setuprow}[2]{#1 & \parbox[t]{0.72\textwidth}{#2}\\[3pt]}
\begin{tabular}{ll}
\hline\hline
Setting & Value or definition \\
\hline
\multicolumn{2}{c}{M87-like runs, code units $G=c=M=1$} \\
\setuprow{Spacetime and EOS}{Kerr--Schild, $a_*=0.5$; ideal gas, $\Gamma=13/9$}
\setuprow{Torus}{Prograde, untilted; $r_{\rm edge}=6$, $r_{\rm max}=12$, $\rho_{\rm max}=1$}
\setuprow{Magnetic seed}{$A_\phi=0.4178\max(\rho-0.2,0)$; fiducial perturbation amplitude zero}
\setuprow{Mesh}{$128\times128\times1$; $r\in[1.7811841173,40]$, $\theta\in[0,\pi]$, $\phi\in[0,2\pi]$}
\setuprow{Mapping}{Uniform $r,\theta$; $h=1$; no refinement}
\setuprow{Density floor}{$\rho\geq\max(10^{-6},10^{-4}r^{-3/2},b^2/100)$}
\setuprow{Pressure floor}{$p\geq\max(10^{-8},10^{-6}r^{-5/2},10^{-3}b^2/2)$}
\setuprow{Velocity ceiling}{$W\leq50$}
\setuprow{Boundaries}{Radial user outflow rules stated in text; polar wedge; periodic azimuth}
\setuprow{Reconstruction}{Piecewise linear ($xorder=2$); HLLE Riemann solver}
\setuprow{Interface treatment}{Direct coordinate-frame fluxes for M87; local interface-frame transformations enabled in the benchmarks}
\setuprow{Integrator}{VL2; localized CFL $0.3$, matched ideal/uniform CFL $0.06022034$; no STS}
\setuprow{Diffusion step}{Equation~(\ref{eq:diffusion_timestep}); matched step $\simeq0.00786$}
\setuprow{Diagnostics}{Flux surface $r_{\rm diag}=2.2$; standard history cadence $0.5$; snapshots $5$}
\setuprow{Custom history}{One row per fluid cycle; common analysis grid spacing $0.5$}
\hline
\multicolumn{2}{c}{Harris matrix, code units $\ell_0=c=1$} \\
\setuprow{Equations and EOS}{GRMHD in Minkowski coordinates; $\Gamma=5/3$}
\setuprow{Mesh and boundaries}{$128\times64$ or $256\times128$, $[-1,1]\times[-0.5,0.5]$; periodic $x$, outflow $y$}
\setuprow{Background}{$\rho_0=B_0=1$, $P_{\rm bg}=0.5$, $a=0.02$; fluid initially at rest}
\setuprow{Perturbation}{Equation~(\ref{eq:harris_az}); $\psi_0=10^{-3}$, $\sigma=0.2$, $m=1$}
\setuprow{Trigger}{Constant controls or Equation~(\ref{eq:trigger}), $\alpha_{\rm crit}=20$, $\delta\alpha=2$}
\setuprow{Time integration}{VL2, piecewise linear reconstruction, HLLE, $C_{\rm CFL}=0.3$; snapshots $0.05$; $t_{\rm end}=3$; no cap}
\setuprow{Harris floors}{Constant density and pressure floors $3.47\times10^{-18}$; $W_{\max}=1000$; no magnetic floors}
\hline
\end{tabular}
\parbox{0.94\textwidth}{$b^2$ is the comoving magnetic-field strength squared.}
\end{table*}

\subsection{Parameter scans}

We carry out two sets of scans, summarized in Table~\ref{tab:runs}:

\textbf{(i)} An $\alphacrit$ scan at fixed $\fcap=0.2$:
$\alphacrit \in \{70,\,85.1,\,100,\,115\}$, plus the ideal G0 baseline
($\etabg=\etaanom=0$).

\textbf{(ii)} An energy-cap scan at fixed $\alphacrit=85.1$:
$\fcap\in\{0.1,\,0.2,\,0.4\}$, plus the same G0 baseline.

In addition we run a fiducial three-case comparison (G0, G1, G3),
where G1 corresponds to a constant background resistivity
$\etabg=10^{-4}$ and zero anomalous part, providing a comparison
with global uniform-diffusivity and uniform-resistivity GRMHD
calculations \citep{Qian2017, Nathanail2025} and with the full
resistive-GRMHD accretion-disk implementation of
\citet[ApJS 244, 10]{2019ApJS..244...10R}.

\subsection{Diagnostics}

\begingroup 
The raw unsigned near-horizon flux and positive-inflow accretion
rate are
\begin{align}
 \PhiBH&=\frac{\sqrt{4\pi}}{2}
  \oint_{r_{\rm diag}}|B^r|\sqrt{-g}\,d\theta\,d\phi,\nonumber\\
 \MdotBH&=-\oint_{r_{\rm diag}}\rho u^r\sqrt{-g}\,d\theta\,d\phi.
 \label{eq:surface_diagnostics}
\end{align}
The diagnostic surface is $r_{\rm diag}=2.2\,\rg$, outside the
horizon at $1.866\,\rg$. We retain the subscript BH for this
common surface. The factor $\sqrt{4\pi}$ converts the native
magnetic units to the customary Gaussian-unit flux normalization.
We form
$\phiBH(t)=\PhiBH(t)/\sqrt{|\MdotBH(t)|\rg^2c}$
at each time before averaging. Under this convention, the
commonly quoted MAD saturation scale is $\phiBH\simeq50$
\citep{Tchekhovskoy2011,Narayan2022}.

The diffusion diagnostic is integrated over the complete grid:
\begin{equation}
 \Qohm(t)=\sum_c q_{{\rm Ohm},c}V_c,\qquad
 V_c=\int_c\sqrt{-g}\,d^3x.
 \label{eq:integrated_q}
\end{equation}
We read the saved cell diagnostic, using the evolution's stored
coefficient and the current evaluated for output. The cap uses
the fluid timestep, independently of the snapshot interval.
Equation~(\ref{eq:integrated_q}) is a coordinate-volume diagnostic,
distinct from a comoving heat source.

For the localized runs, an active cell satisfies
$\eta_c>0.1\eta_{\rm anom}=10^{-4}$. We measure its proper-volume
fraction
\begin{equation}
 f_{\rm act}=
 \frac{\sum_{c\in\mathcal D}\mathbf1_{\eta_c>10^{-4}}\,V_c/\alpha_c}
 {\sum_{c\in\mathcal D}V_c/\alpha_c},
 \label{eq:active_volume}
\end{equation}
where $\mathcal D$ is the inner domain $r\leq10\,\rg$,
$0.08<\theta<\pi-0.08$. The conversion $V_c/\alpha_c$
uses the cell-center lapse.

The standard history cadence is $0.5\,\rg/c$, and the custom
surface diagnostics are recorded every fluid cycle at the completed step time.
Surface histories are linearly interpolated to a common
$0.5\,\rg/c$ grid for trapezoidal time averages.
Snapshot-based activity and density comparisons use a common
$5\,\rg/c$ grid with linear interpolation and trapezoidal
time averages, with identical integration endpoints for both branches.
\endgroup 

\section{Results}\label{sec:results}

\begin{table*}[t]
\caption{M87 comparison matrix.\label{tab:runs}}
\centering
\begin{tabular}{llcccl}
\hline\hline
Model & Prescription & $\alpha_{\rm crit}$ & $f_{\rm cap}$ &
$(\eta_{\rm bg},\eta_{\rm anom})$ & Time span $[\rg/c]$\\
\hline
G0 & ideal & -- & -- & $(0,0)$ & $0$--$500$\\
G1 & uniform & -- & -- & $(10^{-4},0)$ & $0$--$500$\\
G3 & localized & $85.1$ & $0.2$ & $(0,10^{-3})$ & $0$--$1500$\\
Threshold scan & localized & $70,100,115$ & $0.2$ & $(0,10^{-3})$ & $0$--$500$\\
Cap scan & localized & $85.1$ & $0.1,0.4$ & $(0,10^{-3})$ & $0$--$500$\\
Late off & ideal continuation & -- & -- & $(0,0)$ & $1000$--$1500$\\
Late on & localized continuation & $85.1$ & $0.2$ & $(0,10^{-3})$ & $1000$--$1500$\\
\hline
\end{tabular}
\parbox{0.94\textwidth}{The early runs share the initial torus,
grid, and timestep. Both late branches start from the same G3
restart at $t=1000\,\rg/c$. The localized transition width is
$\delta\alpha=8.5$ in every M87 run. The late-on branch also
completes the extended G3 sequence.}
\end{table*}

\subsection{Matched bulk-flow response}
\label{sec:results:G0baseline}
\label{sec:results:global}

\begin{figure*}[t]
\centering
\includegraphics[width=0.78\textwidth]{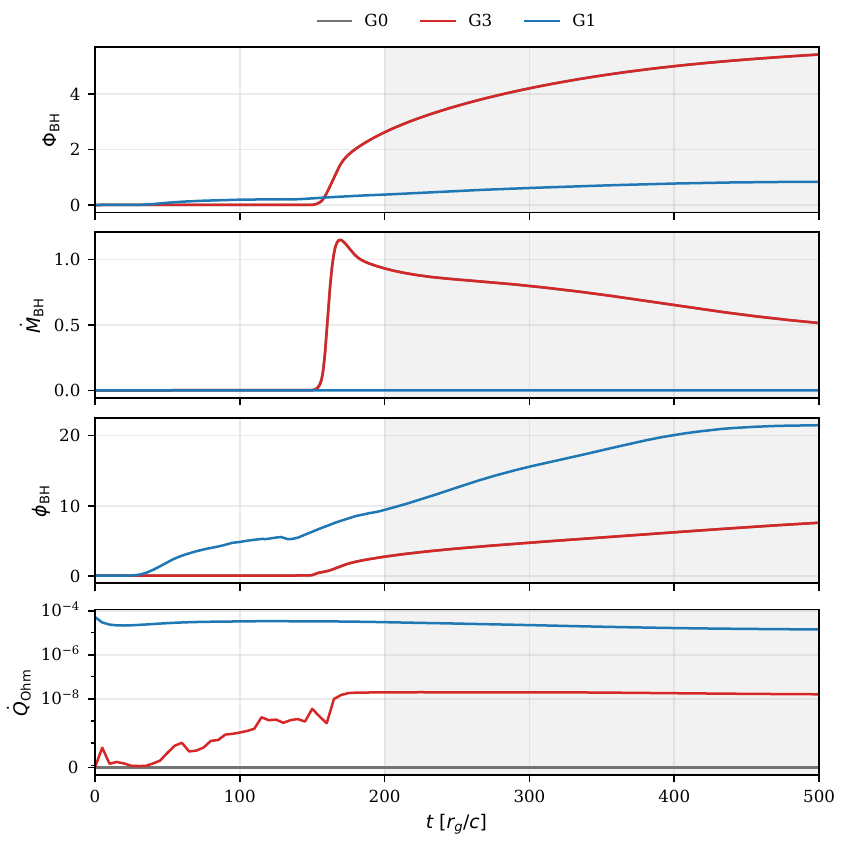}
\caption{Matched ideal (G0), uniform (G1), and localized
(G3) runs. From top to bottom: raw near-horizon flux, accretion
rate, normalized flux, and the full-domain
$\Qohm$ diagnostic. Shading marks the common
$t=200$--$500\,\rg/c$ analysis interval. The bottom panel uses
a symmetric-log scale to display the exactly zero G0 diagnostic
alongside the two nonzero models.}
\label{fig:g0_g1_g3_compare}
\end{figure*}

Figure~\ref{fig:g0_g1_g3_compare} separates the raw magnetic
flux, accretion rate, normalized flux, and diffusion diagnostic.
{Over $t=200$--$500\,\rg/c$, the ideal reference has
$\langle\PhiBH\rangle=4.4705$,
$\langle\MdotBH\rangle=0.71932$, and
$\langle\phiBH\rangle=5.4009$. The normalized flux has a
temporal standard deviation of $1.34$, identifying a sub-MAD
reference state under our stated normalization.}

For a common time interval, we define the percentage shift
$\delta_X=100(\langle X\rangle/\langle X\rangle_{\rm ref}-1)$.
The radial-density distance is
\begin{equation}
 D_\rho=
 \left[
 \frac{\sum_{3<r/\rg<20}
  (\bar\rho_{\rm model}-\bar\rho_{\rm ref})^2\Delta r}
 {\sum_{3<r/\rg<20}\bar\rho_{\rm ref}^{\,2}\Delta r}
 \right]^{1/2}.
 \label{eq:density_l2}
\end{equation}
Here $\bar\rho$ is averaged over the stated snapshot interval
and over $|\theta-\pi/2|\leq\pi/3$ with
$\sin\theta\,d\theta$ weighting.
{This is a finite-time matched structural diagnostic for an evolving
flow. It does not assume inflow equilibrium throughout
$3<r/\rg<20$.}

{The localized G3 mean shifts are below $0.02\%$ for all three
surface diagnostics, with $D_\rho=0.005\%$
(Table~\ref{tab:global_preservation}). The ideal time variations
are much larger: their standard deviations are $17.3\%$,
$16.6\%$, and $24.9\%$ of the raw-flux, accretion-rate, and
normalized-flux means.} G0 and G3 therefore retain closely
matching bulk solutions while G3 supplies a nonzero
current-sheet diagnostic.

{At the stated uniform coefficient, G1 reduces the mean raw
flux by $85.0\%$ and the accretion rate by $99.8\%$, with
$D_\rho=35.4\%$. Its normalized flux increases to $17.17$}
because the accretion-rate denominator falls more rapidly than
the raw flux. At this coefficient, G1 illustrates
the global effect of broadly distributed diffusion.

\begin{table*}[t]
\caption{{Global diagnostics and percentage shifts in matched
comparisons.}\label{tab:global_preservation}}
\centering
\setlength{\tabcolsep}{4pt}
\begin{tabular}{lrrrrrrr}
\hline\hline
Model & $\langle\PhiBH\rangle$ & $\delta_\Phi$ &
$\langle\MdotBH\rangle$ & $\delta_{\dot M}$ &
$\langle\phiBH\rangle$ & $\delta_\phi$ & $100D_\rho$\\
\hline
\multicolumn{8}{c}{Early comparison: $t=200$--$500\,\rg/c$}\\
G0 & $4.47047$ & $0$ & $0.719317$ & $0$ & $5.40087$ & $0$ & $0$\\
G3 & $4.47047$ & $-0.000055$ & $0.719436$ & $+0.0165$ & $5.40019$ & $-0.0126$ & $0.00496$\\
G1 & $0.67197$ & $-84.969$ & $0.001541$ & $-99.786$ & $17.1746$ & $+217.996$ & $35.365$\\
\hline
\multicolumn{8}{c}{Late common-restart comparison: $t=1050$--$1500\,\rg/c$}\\
Off & $4.30825$ & $0$ & $0.864693$ & $0$ & $4.95385$ & $0$ & $0$\\
On & $4.26291$ & $-1.0525$ & $0.851202$ & $-1.5603$ & $4.92564$ & $-0.5694$ & $0.294$\\
\hline
\end{tabular}
\parbox{0.94\textwidth}{All shifts and the final column are
percentages. The reference is G0 in the early comparison and
the off continuation in the late comparison. Means use the
common $0.5\,\rg/c$ history grid.
The normalized flux is computed pointwise before averaging;
$D_\rho$ follows Equation~(\ref{eq:density_l2}).
Means and percentage shifts are computed before rounding.}
\end{table*}

\begin{figure*}[t]
\centering
\includegraphics[width=0.98\textwidth]{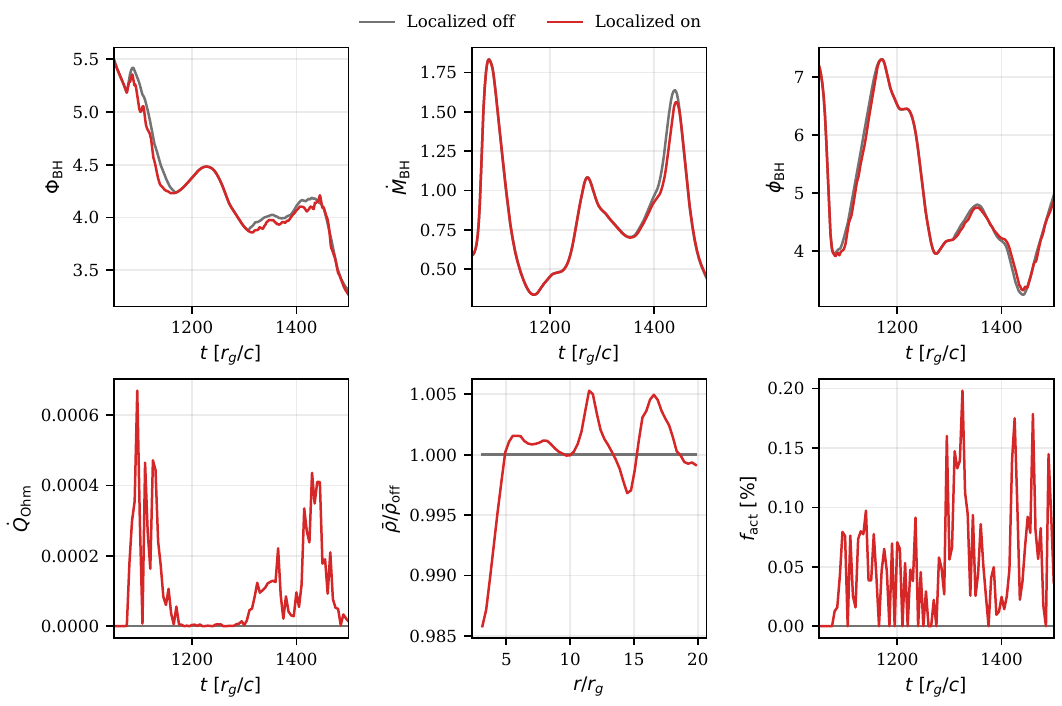}
\caption{{Late matched on/off comparison from the common
$t=1000\,\rg/c$ restart.} The upper panels show the three
surface diagnostics over $t=1050$--$1500\,\rg/c$. The lower
panels show $\Qohm$, the time-averaged density-profile ratio,
and the inner proper-volume active fraction. The localized-on
branch develops intermittent activity while following the
off-branch bulk evolution.}
\label{fig:late_matched}
\end{figure*}

\begingroup 
The common-restart experiment (Figure~\ref{fig:late_matched}) tests
bulk-flow preservation during the late active phase.
Both branches start at $t=1000\,\rg/c$
from the same magnetic and fluid state and use the same timestep.
Over $t=1050$--$1500\,\rg/c$, the localized-on changes in
$\PhiBH$, $\MdotBH$, and $\phiBH$ are $-1.05\%$,
$-1.56\%$, and $-0.57\%$, with $D_\rho=0.29\%$.
The corresponding off-branch fluctuations are $11.5\%$,
$44.3\%$, and $23.1\%$ of their means. Each mean shift is
less than $0.10$ times the reference temporal standard deviation.
\endgroup 

\subsection{Spatial support and magnetic structure}

\begin{figure*}[t]
\centering
\includegraphics[width=0.98\textwidth]{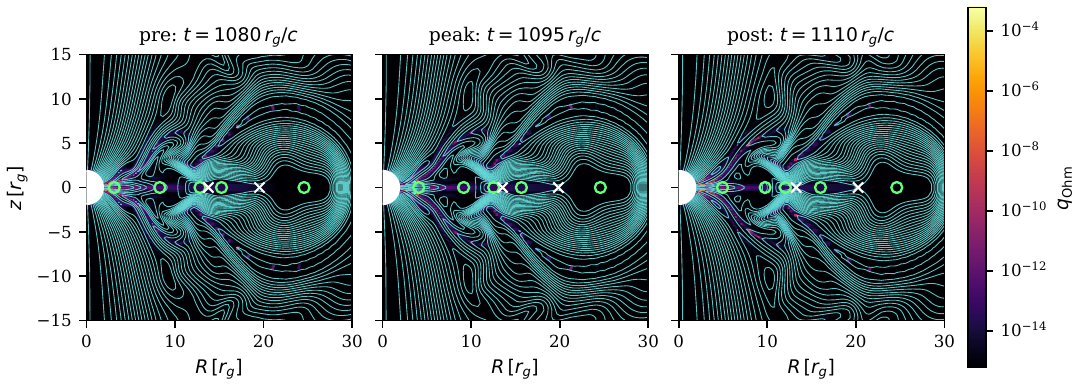}
\caption{Native $\qohm$ and poloidal flux contours at
{$t=1080$, $1095$, and $1110\,\rg/c$}, before, at, and after the
sampled heating peak, shown in $R=r\sin\theta$ and $z=r\cos\theta$.
All panels share the logarithmic color scale and the $A_\phi$ contour levels.
{Cyan contours show $A_\phi$; open green circles and white crosses mark
near-equatorial O-points and X-points, respectively.}
The white disk marks the event horizon at $r_+=1.866\,\rg$.}
\label{fig:qohm_map}
\end{figure*}

Figure~\ref{fig:qohm_map} shows the native $\qohm$ map around
the strongest sampled late event, at {$t=1095\,\rg/c$.
The full-domain diagnostic reaches $6.69\times10^{-4}$.
At this peak, approximately $90\%$ of its integral lies at
$r<3\,\rg$, and more than $99.9\%$ lies within
$|\theta-\pi/2|<0.3$. The selected region is an inner
equatorial current layer. The inner-domain active fraction
averages $0.053\%$ over $t=1050$--$1500\,\rg/c$ and reaches
$0.20\%$.}

\begingroup 
Axisymmetry also permits the reconstruction
\begin{equation}
 \partial_\theta A_\phi=\sqrt{-g}\,B^r,\qquad
 \partial_r A_\phi=-\sqrt{-g}\,B^\theta.
 \label{eq:aphi_reconstruction}
\end{equation}
\endgroup 
We integrate the first relation from the regular polar axis and
use the second as a cross-component check. Across the three
displayed snapshots, the reconstructed and saved polar
components have correlations above $0.998$ and normalized RMS
differences of $0.060$--$0.065$ for $3<r/\rg<30$.
Closed contours identify magnetic islands. Equatorial candidates
seed a two-dimensional stationary-point search; the roots are
classified from the Hessian of $A_\phi$. {Closed
structures are present farther along the equatorial field,
whereas the largest diffusion diagnostic is concentrated in the
inner current layer.}

\subsection{Late activity and raw-flux evolution}
\label{sec:results:longrun}
\label{sec:results:erupt_align}

\begin{figure*}[t]
\centering
\includegraphics[width=0.92\textwidth]{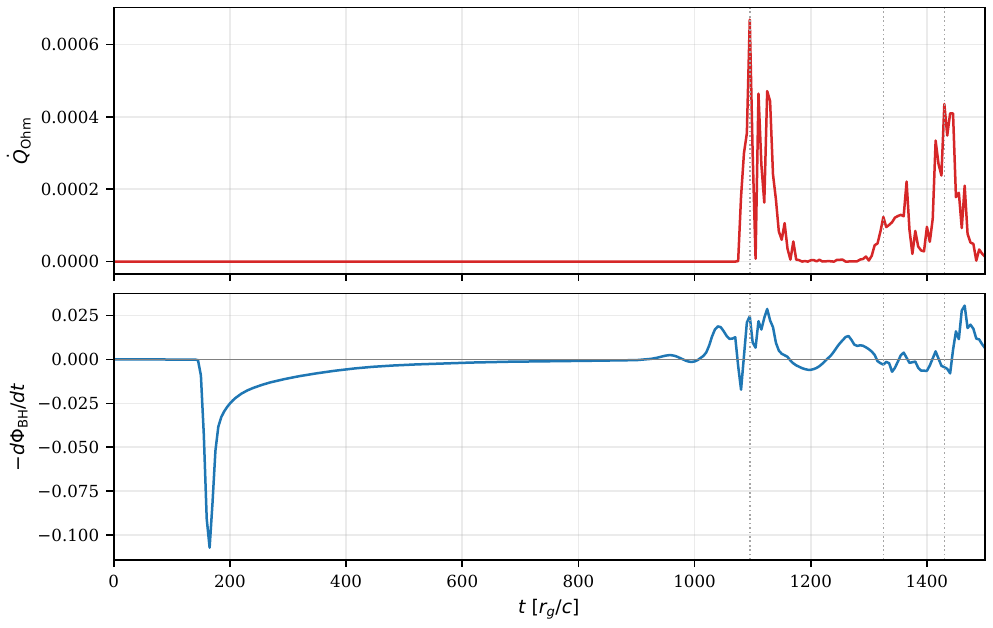}
\caption{Extended G3 activity and raw-flux evolution.
The upper panel shows the full-domain $\Qohm$ diagnostic,
and the lower panel shows $\Sphi=-d\PhiBH/dt$ using the
common $5\,\rg/c$ grid. Vertical lines mark the three
selected heating events. The zero line distinguishes raw-flux
loss from recovery.}
\label{fig:longrun}
\end{figure*}

In the extended G3 sequence (Figure~\ref{fig:longrun}),
{the mean $\Qohm$ increases from
$1.87\times10^{-8}$ in the early comparison to
$1.08\times10^{-4}$ in the late comparison.
A peak finder with prominence $10^{-5}$ and minimum spacing
$100\,\rg/c$ selects three late events at
$t\simeq1095$, $1325$, and $1430\,\rg/c$.
Their sampled peaks are $6.69\times10^{-4}$,
$1.23\times10^{-4}$, and $4.36\times10^{-4}$.}

\begingroup 
We define $\Sphi=-d\PhiBH/dt$, positive when the raw flux
decreases, and
$r_{QS}=\operatorname{corr}[\Qohm(t),\Sphi(t)]$.
Both signals are sampled on a uniform $5\,\rg/c$ grid by
linear interpolation; $\Sphi$ uses centered differences in the
interior and one-sided differences at the endpoints, with no
additional smoothing. In $\pm50\,\rg/c$ event windows, the
zero-lag correlations $r_{QS}$ are $0.45$, $-0.63$, and $-0.19$.
\endgroup 
The first peak accompanies decreasing raw flux; the other two
occur during intervals of flux recovery. Their different signs
show that recurrent local activity is not tied to a single
phase of the global flux evolution. These associations do not establish
a causal ordering.

\subsection{Threshold and cap dependence}
\label{sec:results:robust}
\label{sec:results:cap}

\begin{figure*}[tp]
\centering
\includegraphics[width=0.84\textwidth]{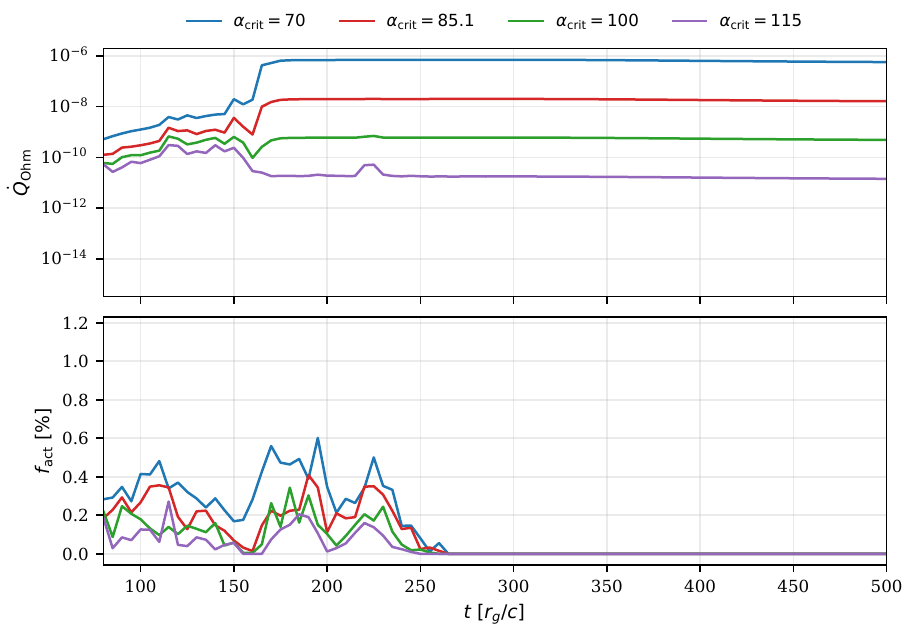}
\caption{Threshold dependence at fixed
$\eta_{\rm anom}=10^{-3}$ and $f_{\rm cap}=0.2$.
Top: full-domain $\Qohm$.
Bottom: the inner proper-volume active fraction of
Equation~(\ref{eq:active_volume}).
The displayed and averaged interval is $80$--$500\,\rg/c$.}
\label{fig:alpha_scan}
\end{figure*}

The fixed-threshold scan measures how spatial selection changes
the diffusion diagnostic. Figure~\ref{fig:alpha_scan} and
Table~\ref{tab:results} cover
{$80\leq t/(\rg/c)\leq500$}. Increasing $\alpha_{\rm crit}$
reduces the diagnostic amplitude and the proper-volume
occupation. The tanh tail also contributes below the active-cell threshold.

\begin{table*}[t]
\caption{{Activity and bulk response in the
threshold and cap scans.}\label{tab:results}}
\centering
\setlength{\tabcolsep}{5pt}
\begin{tabular}{lrrrrr}
\hline\hline
Model & $\langle\Qohm\rangle$ & $\max\Qohm$ &
$\langle f_{\rm act}\rangle$ [\%] & $\max f_{\rm act}$ [\%] &
$\delta_{\dot M}$ [\%]\\
\hline
G0 & $0$ & $0$ & $0$ & $0$ & $0$\\
G3-70 & $5.24\times10^{-7}$ & $7.00\times10^{-7}$ & $0.138$ & $0.601$ & $+0.0215$\\
G3-85 & $1.51\times10^{-8}$ & $2.03\times10^{-8}$ & $0.0873$ & $0.409$ & $+0.0165$\\
G3-100 & $5.13\times10^{-10}$ & $6.97\times10^{-10}$ & $0.0547$ & $0.344$ & $+0.0139$\\
G3-115 & $4.17\times10^{-11}$ & $3.04\times10^{-10}$ & $0.0328$ & $0.271$ & $+0.0116$\\
G3, cap $0.1$ & $1.51\times10^{-8}$ & $2.03\times10^{-8}$ & $0.0873$ & $0.409$ & $+0.0165$\\
G3, cap $0.4$ & $1.51\times10^{-8}$ & $2.03\times10^{-8}$ & $0.0873$ & $0.409$ & $+0.0165$\\
\hline
\end{tabular}
\parbox{0.94\textwidth}{{The heating and activity columns use
$80\leq t/(\rg/c)\leq500$ and the common $5\,\rg/c$ snapshot
grid. The final column is the $200$--$500\,\rg/c$ accretion-rate
shift relative to G0, matching the bulk-comparison window.
The activity fraction is proper-volume weighted in the inner
domain.}}
\end{table*}

{Across the four thresholds, the early accretion-rate shift
remains below $0.022\%$ and the normalized-flux shift below
$0.018\%$. Varying $f_{\rm cap}$ from $0.1$ to $0.4$
gives the same bulk and activity metrics at the table's precision;
the mean $\Qohm$ differs by less than $0.003\%$.}

\subsection{Ideal-flow resolution context}
\label{sec:results:convergence}

\begin{figure*}[tp]
\centering
\includegraphics[width=0.84\textwidth]{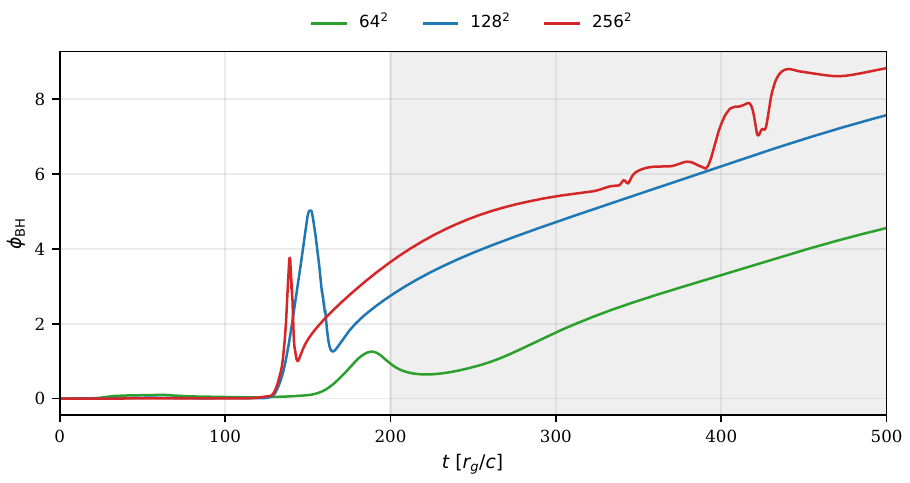}
\caption{Independent ideal-flow resolution sequence at
$C_{\rm CFL}=0.3$. Shading marks the $200$--$500\,\rg/c$ averaging
window. This sequence gives a resolution scale for the absolute
flux. The principal G0/G3 comparison uses the smaller common
timestep specified in Table~\ref{tab:numerical_setup}.}
\label{fig:convergence}
\end{figure*}

An independent ideal-GRMHD sequence at $C_{\rm CFL}=0.3$
provides the resolution context
(Figure~\ref{fig:convergence}). Its mean normalized fluxes over
$t=200$--$500\,\rg/c$ are $2.51$, $5.41$, and $6.36$ on
$64^2$, $128^2$, and $256^2$ grids, with temporal standard
deviations $1.27$, $1.34$, and $1.58$.
The fine-grid value is about $18\%$ above the intermediate-grid
value. The prescription response is measured separately with
matched grids and timesteps: the early G3--G0 normalized-flux
shift is $-0.013\%$, far smaller than this resolution scale.

\section{Discussion}\label{sec:discussion}

\subsection{Spatial selection and bulk-flow preservation}

Ideal-GRMHD calculations characterize candidate current sheets
\citep{Ball2018}, and sufficiently resolved MAD simulations
follow plasmoid-mediated reconnection \citep{Ripperda2022}.
Uniform-resistivity calculations show how non-ideal transport
can change global flux accumulation \citep{Qian2017,Nathanail2025}.
Our comparison addresses a complementary question: how much
does a spatially selected diffusion term change an existing
accretion solution? The early G0/G3 match and the late
common-restart experiment answer this question at fixed grid and
timestep. The localized term produces a measurable current-sheet
diagnostic while its mean bulk response remains below the ideal
flow's temporal fluctuations. The uniform control demonstrates
the contrasting response at the stated broadly supported
coefficient.

The Fourier, traveling-wave, and Schwarzschild energy tests
verify that the conservative update transfers magnetic energy to the
gas while accounting for compression and boundary transport.

\subsection{Interpretation of the current-sheet diagnostic}

The empirical trigger supplies a reproducible control on
where diffusion acts. Its threshold changes both the low-amplitude
tanh tail and the fraction of cells with appreciable diffusivity.
PIC-derived effective-resistivity models provide a route toward
physically recalibrating or replacing this trigger in matched
current-sheet geometries \citep{Moran2025,Ripperda2026}.

The Harris benchmarks establish consistency of the local-rate
exponent with $1/2$, rather than uniquely identifying a steady
Sweet--Parker regime (Section~\ref{sec:vandv}).

\subsection{Use in radiative models}

The saved $\qohm(\mathbf x,t)$ field supplies a spatial and
temporal weight for dissipative regions in an accretion solution
used for synthetic observations \citep{EHT2019PaperV}.
The small bulk-flow response in the matched comparisons
provides a controlled setting for comparing dissipation-weighted
emission models.

{Converting the coordinate-curl diagnostic $\eta J^2$ into a
comoving electron-heating rate requires a specified frame
conversion and an electron--ion partition,
$Q_e=f_eQ_{\rm diss}$. A separate electron entropy or internal-energy
equation can then incorporate that source, together with the
chosen cooling, conduction, and Coulomb exchange.}
The formulation of \citet{2015MNRAS.454.1848R} illustrates this
additional closure. The diagnostic can also be compared with
phenomenological temperature prescriptions such as the
R--$\beta$ model \citep{Moscibrodzka2016}; it does not by itself
set the electron temperature.

\subsection{Dynamical setting}

Unlike full Maxwell--Ohm formulations
\citep{Komissarov2007,Palenzuela2009,Mignone2019,2019ApJS..244...10R},
the present coordinate-frame prescription does not evolve an
independent electric field. The accretion-flow comparisons concern
axisymmetric, sub-MAD flows over finite time intervals.

\section{Conclusions}\label{sec:conclusions}

We present a spatially selective magnetic-diffusion prescription
with explicit energy bookkeeping in \Athena{}. A smooth
current-based trigger sets the coefficient; CT updates the field,
and the conservative energy-flux correction determines thermal
transfer through primitive recovery. There is no additional
$\eta J^2$ energy source.

The validation measures evolved gas energy as well as magnetic
loss. Periodic Fourier and traveling Alfv\'en tests have no
floor repairs and close the total-energy budget to roundoff.
A divergence-free Schwarzschild test verifies the geometric
construction and open-boundary Killing-energy balance.
The Harris matrix yields localized exponents
$0.510\pm0.027$ and $0.488\pm0.028$, consistent with $1/2$
over $t=0.5$--$2.0$; the same-resolution
$\eta=10^{-3}$ diagnostic exceeds the ideal reference by 5.96.

Matched M87-like experiments quantify the dynamical effect.
The early surface-mean shifts remain below $0.02\%$, with
$D_\rho=0.005\%$. During the late active phase, the common-restart
on/off shifts remain below $1.6\%$, with $D_\rho=0.29\%$,
while $\Qohm$ reaches $6.69\times10^{-4}$.
These results demonstrate measurable, spatially concentrated
diffusion with a small bulk-flow response under the tested
conditions.

\begin{acknowledgments}
We acknowledge support from the start-up funding of
Zhejiang University and the Zhejiang Provincial Top-Level Research
Support Program. The simulations and analysis presented in this
article were carried out on the SilkRiver Supercomputer of Zhejiang
University. We acknowledge the developers of \Athena{}
\citep{Stone2020}.
We used ChatGPT and Codex (OpenAI) to assist with manuscript editing,
code development, and data analysis. The authors take full
responsibility for the content of this paper.
\end{acknowledgments}

\software{\Athena{} \citep{Stone2020},
NumPy \citep{numpy2020},
SciPy \citep{scipy2020},
matplotlib \citep{matplotlib2007}}

% Keep main-text figures and tables out of the appendix.
\onecolumngrid
\clearpage
\appendix
\renewcommand{\theHequation}{A.\arabic{equation}}

\section{Additional Verification Tests}
\label{sec:appendix_vnv}

\subsection{Cartesian Flat-Space Diffusion}

\begin{figure*}[t]
\centering
\includegraphics[width=0.90\textwidth]{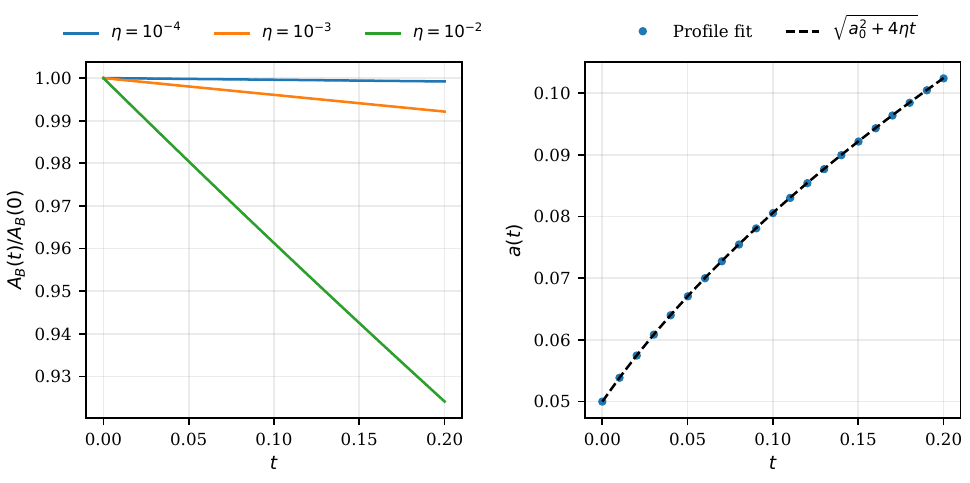}
\caption{Cartesian diffusion profiles. Left: normalized
Fourier amplitude for three resistivities; dashed curves are
$\exp(-\eta k^2t)$. {Right: the fitted error-function scale
$a(t)$ and the analytic $\sqrt{a_0^2+4\eta t}$.
Both comparisons use 256 cells and $t\leq0.2$.}}
\label{fig:cartesian_multi_eta}
\end{figure*}

The linear-amplitude Fourier test uses
$B_y=10^{-6}\sin(2\pi x)$, $\rho=p=1$, $\Gamma=5/3$,
and initial $v=0$ on a periodic unit interval with 256 cells.
We use VL2, PLM, HLLE, $C_{\rm CFL}=0.3$, and
$\eta\in\{10^{-4},10^{-3},10^{-2}\}$, evolving to $t=0.2$.
The diffusion equation has amplitude decay rate $\eta k^2$.
{A least-squares fit to $\ln A_B(t)$ gives relative rate errors
$0.393\%$, $0.0178\%$, and $0.0166\%$, respectively
(Figure~\ref{fig:cartesian_multi_eta}).}

\subsection{Dynamic Energy Transfer and Sheet Diffusion}
\label{sec:wave_ledger}

\textbf{A2.} On the periodic unit interval, a weak right-moving
Alfv\'en wave has
\begin{align}
 v^y&=-v_0\sin(2\pi x), &v_0&=0.01,\nonumber\\
 B^y&=\sqrt{\rho_0h_0+(B^x)^2}\,v_0\sin(2\pi x),
 &B^x&=0.5.
\end{align}
The background is $\rho_0=p_0=1$, $\Gamma=5/3$, and the primitive
velocity is initialized as $\tilde u^i=Wv^i$. Both the $\eta=0.01$
and ideal controls use 256 cells, VL2, PLM, HLLE, and
$C_{\rm CFL}=0.3$, and are evolved to $t=2$. \begingroup The volume-integrated
components of the relativistic conserved energy are
\begin{align}
 E_{\rm rest}&=\int\rho W\,dV,&
 E_{\rm kin}&=\int\rho W(W-1)\,dV,\nonumber\\
 E_{\rm th}&=\int\!\left[\frac{\Gamma p W^2}{\Gamma-1}-p\right]dV,\nonumber\\
 E_B&=\frac12\int B^2dV,&
 E_E&=\frac12\int|\boldsymbol v\times\boldsymbol B|^2dV.
\end{align}
Their sum reproduces $-\int T^0{}_0dV$. In particular,
$E_{\rm th}$ differs from the comoving internal-energy density
integrated on the coordinate slice, $E_{\rm int}=\int p/(\Gamma-1)dV$.
For the uniform initial entropy $K_0=p_0/\rho_0^\Gamma$, the
irreversible thermal excess is measured as
\begin{equation}
 E_{\rm irr}=\int\frac{p-K_0\rho^\Gamma}{\Gamma-1}\,dV.
 \label{eq:entropy_excess}
\end{equation}
\endgroup 
{At $t=2$, the magnetic and kinetic changes are
$-5.1992\times10^{-5}$ and $-1.3429\times10^{-5}$, respectively;
the lab-frame thermal contribution increases by
$6.8777\times10^{-5}$.} Equation~(\ref{eq:entropy_excess}) gives
$1.02344\times10^{-4}$, agreeing with $I_Q$ to $0.0080\%$.

\textbf{A3.} The initializer uses
$B_y(x,0)=B_0\operatorname{erf}[(x-x_0)/a_0]$. \begingroup Its diffusion solution is
\begin{equation}
 B_y(x,t)=B_0\operatorname{erf}\!\left[
 \frac{x-x_0}{\sqrt{a_0^2+4\eta t}}\right].
 \label{eq:erf_solution}
\end{equation}
\endgroup 
We use $B_0=10^{-4}$, $a_0=0.05$, $x_0=0.5$,
$\rho=p=1$, $\Gamma=5/3$, and $\eta=0.01$ on $0<x<1$
with 256 cells and outflow boundaries. The integrator, solver,
and CFL are the same as in A1. {A one-parameter fit of the
saved $B_y/B_0$ profile gives a maximum relative width error
of $0.072\%$ through $t=0.2$. At that time,
$\langle|B_y-B_{y,\rm analytic}|\rangle/B_0=7.89\times10^{-5}$.}

\subsection{Schwarzschild Initial Constraint (A5)}
\label{sec:a5}
\begingroup 
We initialize line-integrated vector-potential values on CT edges.
For integer $m\geq1$, let $A_r=A_\theta=0$
and
\begin{align}
 A_\phi&=A_0 f(r)\sin^{m+2}\theta\cos(m\phi),\nonumber\\
 f(r)&=\sin^2\!\left[\pi\frac{r-r_1}{r_2-r_1}\right],
 \quad r_1\leq r\leq r_2.
\end{align}
Set $f=0$ outside this shell. The nonzero coordinate fields are
\begin{equation}
 B^r=\frac{\partial_\theta A_\phi}{\sqrt{-g}},\qquad
 B^\theta=-\frac{\partial_r A_\phi}{\sqrt{-g}}.
\end{equation}
Their face fluxes are the oriented sums of edge integrals, ensuring
that the discrete divergence of the discrete curl cancels.
The azimuthal edge integral uses
$[\sin(m\phi_+)-\sin(m\phi_-)]/m$ exactly. We set
$A_0=0.1$, $m=2$, $r_1=8$, and $r_2=10$.
\endgroup 
{
The full computational domain is $r\in[8,10]$,
$\theta\in[\pi/4,3\pi/4]$, and $\phi\in[0,2\pi]$
in Athena++ Schwarzschild coordinates with $M=1$.
The $32\times16\times32$ grid has uniform coordinate-face
spacing in all three directions.}
Radial and polar boundaries are outflow; the azimuth is periodic.
The fluid begins at rest with $\rho=1$, $p=0.1$, and $\Gamma=5/3$.
We evolve $\eta=0.01$ to $t=0.2$ with VL2, HLLE, PLM, CFL $0.3$,
and Equation~(\ref{eq:diffusion_timestep}); the timestep is
$6.56\times10^{-4}$. This test allows both gravitational motion
and open-boundary energy transport.

\begingroup 
The constraint diagnostic is the native CT flux imbalance,
\begin{equation}
 \begin{aligned}
 D_{B,c}&=\frac{1}{V_c}\sum_{f\in\partial c}s_{cf}A_fB_f,\\
 \epsilon_{B,c}&=\frac{|D_{B,c}|V_c}
 {\max\!\left(|B_c|\sum_f A_f,10^{-30}\right)}.
 \end{aligned}
\end{equation}
Here $|B_c|=\sqrt{\gamma_{ij}B_c^iB_c^j}$ is the norm of the
native cell-centered coordinate field. The denominator lower
bound is in code units. Both constraint maxima include all
interior cells, including zero-field cells.
We obtain $\|D_B(0)\|_\infty=5.98\times10^{-18}$,
$\max_t\|D_B\|_\infty=1.35\times10^{-16}$, and
$\max_{t,c}\epsilon_{B,c}=1.56\times10^{-15}$.
The Killing energy increases by $0.01353174$, balanced by a
net boundary influx of the same magnitude. No floor repair occurs;
the largest residual is $2.98\times10^{-12}$ in absolute units,
or $1.79\times10^{-15}$ of the initial energy.
\endgroup 

\subsection{Fourier Energy-Transfer Resolution Check}

\textbf{A4.} Table~\ref{tab:energy_resolution} reports the recovered thermal
gain, independently accumulated $I_Q$, and their relative difference
in the finite-amplitude Fourier test. All four grids use
the initial conditions and final time of the finite-amplitude
Fourier test in Section~\ref{sec:vandv}.
\begin{table}[t]
\centering
\caption{{Fourier energy ledger at $t=0.2$.}\label{tab:energy_resolution}}
\begin{tabular}{rrrr}
\hline\hline
$N_x$ & $10^5\Delta E_{\rm int}$ & $10^5I_Q$ &
$|\Delta E_{\rm int}/I_Q-1|$ [\%]\\
\hline
64  & 9.12556 & 9.09370 & 0.3504\\
128 & 9.12917 & 9.12128 & 0.0864\\
256 & 9.13218 & 9.12729 & 0.0535\\
512 & 9.13319 & 9.12868 & 0.0494\\
\hline
\end{tabular}
\end{table}
{No floor repairs occur. The total-energy residual remains below
$5.4\times10^{-16}$ relative to the initial energy.}
The thermal-diagnostic difference includes the finite-amplitude fluid response.

\subsection{Resistive Shock Tube (Brio--Wu)}

\begin{figure*}[t]
\centering
\includegraphics[width=0.80\textwidth]{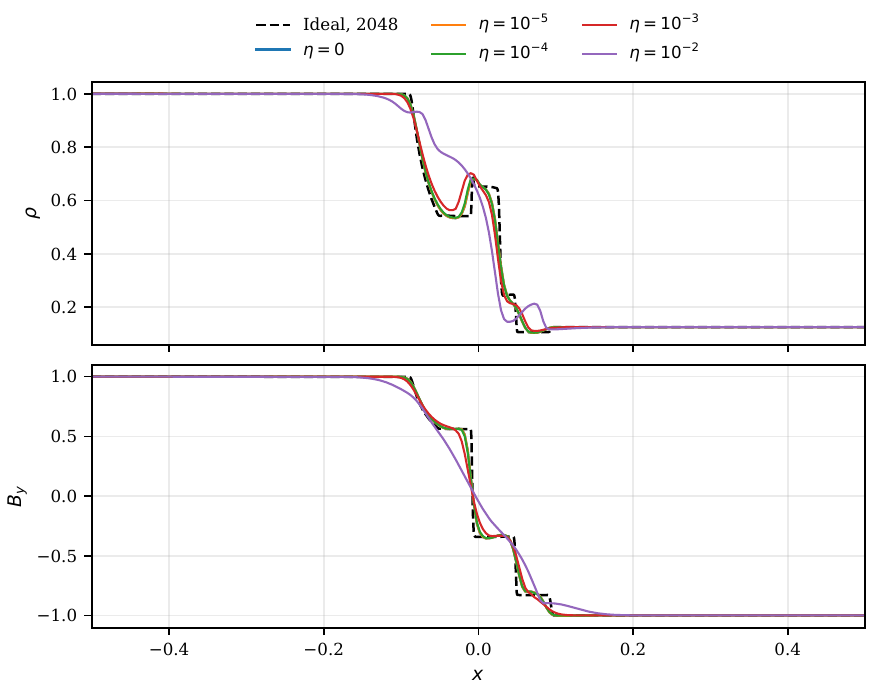}
\caption{Relativistic Brio--Wu-like shock sequence.
Top: density. Bottom: transverse magnetic field.
The five 256-cell solutions are compared with the 2048-cell
ideal reference (black dashed); the $\eta=0$ and $10^{-5}$
curves nearly coincide.}
\label{fig:shock_tube_scan}
\end{figure*}

We run a Brio--Wu-like shock tube at
$\eta\in\{0,10^{-5},10^{-4},10^{-3},10^{-2}\}$ on 256 cells,
evolved to $t=0.1$ (Figure~\ref{fig:shock_tube_scan}).
{The equations are GRMHD in Minkowski coordinates with
$\Gamma=2$.} On $-0.5<x<0.5$, the left and right states have
$(\rho,p,B_y)=(1,1,1)$ and $(0.125,0.1,-1)$,
$B_x=0.75$, and zero initial velocity and $B_z$. Both boundaries
are outflow. The runs use PLM, HLLE, VL2, and $C_{\rm CFL}=0.4$,
with the same constant floor and velocity-ceiling defaults
as the Harris tests.
{The mean absolute density differences from the interpolated
2048-cell ideal reference are $0.00551$ at $\eta=0$ and $0.00555$
at $10^{-5}$; the corresponding $B_y$ differences are
$0.00928$ and $0.00931$.} The small-$\eta$ solution approaches
the same-resolution ideal solution, and increasing $\eta$
broadens the magnetic transitions.

\subsection{Harris Current Sheet}
\label{sec:harris_appendix}

We run a Harris current sheet at $\eta = 10^{-3}$ on a
$128 \times 64$ grid using the GRMHD module in Minkowski
coordinates and $\Gamma=5/3$. Times are measured in $\ell_0/c$,
with $\ell_0=1$. The pressure-balanced
background at rest is
\begin{align}
B_{x,0}(y) &= B_0\tanh(y/a), & B_{y,0}&=B_{z,0}=0,\\
\rho_0(y) &= \rho_0, &
P_0(y)&=P_{\rm bg}+\frac{B_0^2}{2}\operatorname{sech}^2(y/a).
\end{align}
The actual numerical initial state includes a divergence-free
perturbation generated from
\begin{equation}
A_z(x,y)=B_0a\ln\!\cosh(y/a)
+\psi_0\cos[k_x(x-x_{\rm mid})]\exp[-(y/\sigma)^2],
\label{eq:harris_az}
\end{equation}
where $k_x=2\pi m/L_x$. Its continuous magnetic components are
\begin{align}
B_x &= B_0\tanh(y/a)
-\frac{2y\psi_0}{\sigma^2}\cos[k_x(x-x_{\rm mid})]
  e^{-(y/\sigma)^2},\\
B_y &= k_x\psi_0\sin[k_x(x-x_{\rm mid})]e^{-(y/\sigma)^2}.
\end{align}
The code initializes the face-centered field with the discrete curl
of Equation~(\ref{eq:harris_az}), preserving the CT divergence
constraint. We use $B_0=1$, $a=0.02$, $\rho_0=1$,
$P_{\rm bg}=0.5$, $\psi_0=10^{-3}$, $\sigma=0.2$, and $m=1$.

\begin{figure}[!htbp]
\centering
\includegraphics[width=0.95\columnwidth]{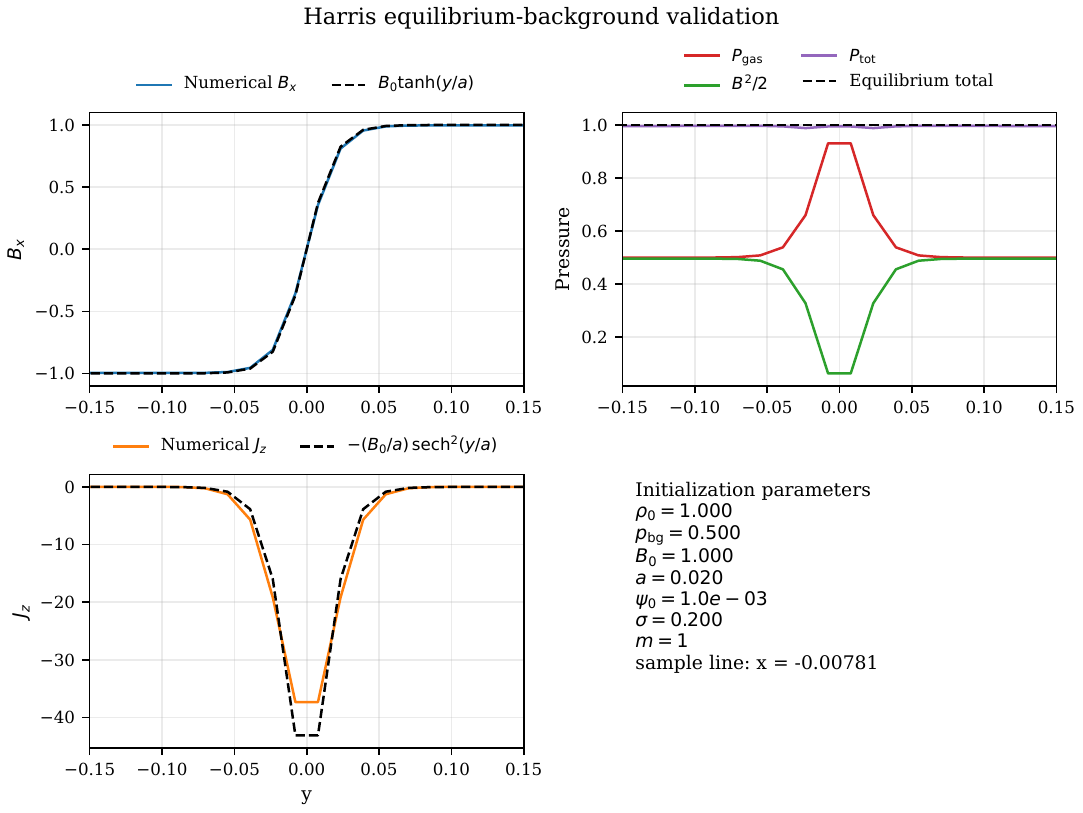}
\caption{Harris equilibrium-background validation for the complete
perturbed initial state in Equation~(\ref{eq:harris_az}).
\emph{Top left:} numerical $B_x(y)$ (solid) and the unperturbed
$B_0\tanh(y/a)$ profile (dashed). \emph{Top right:} thermal,
magnetic, and total pressure, with the unperturbed constant total
pressure shown dashed. \emph{Bottom left:} numerical $J_z$ and the
unperturbed analytic current. \emph{Bottom right:} initialization
parameters and sample-line position.}
\label{fig:harris_ic}
\end{figure}

Figure~\ref{fig:harris_ic} validates the equilibrium background on
the grid. The reconnecting field and current follow the analytic
$\tanh$ and $-(B_0/a)\operatorname{sech}^2(y/a)$ profiles. The
finite perturbation and discrete representation produce a maximum
$1.2\%$ total-pressure deviation on the displayed sample line.

In addition to this structural run, the scaling matrix contains
constant and localized models using the linear trigger
variable $\alpha_J$ in Equation~(\ref{eq:trigger}) at
$128\times64$ and $256\times128$, with four nominal
resistivities per model and resolution.  Symmetry fixes the
primary X-point at the geometric center throughout the scaling analysis.
For the localized cases, the trigger is saturated at the primary
X-point and $\eta_{\rm eff}(X)/\eta_{\rm nominal}=1$ within
saved-output precision. The standard local rate is normalized by the
time-independent initial upstream state at $|y|=10a$.

The common analysis interval ends at $t=2$, before the
high-resolution rate develops its late rapid rise.

\begingroup 
Table~\ref{tab:harris_windows} varies both endpoints using the
same sixteen cases and their $0.05\,\ell_0/c$ snapshot
cadence. Each fit contains four case means. Its quoted error is the
regression scatter across those means, not a confidence interval
obtained by treating every time sample as independent. The local
model's exponents in these early windows remain close to $1/2$.
Extending the same sixteen-case analysis through $t=3$ changes
the localized exponents to $0.449\pm0.019$ and
$0.123\pm0.025$; the later high-resolution response strongly
changes the longer-window fit.
\endgroup 

\begin{table*}[t]
\caption{{Harris local-rate exponent versus analysis window.}
\label{tab:harris_windows}}
\centering
\begin{tabular}{lcccc}
\hline\hline
$[t_{\min},t_{\max}]$ &
Constant, $128\times64$ & Constant, $256\times128$ &
Localized, $128\times64$ & Localized, $256\times128$\\
\hline
$[0.5,1.5]$  & $0.616\pm0.015$ & $0.578\pm0.015$ & $0.531\pm0.026$ & $0.503\pm0.030$ \\
$[0.5,2.0]$  & $0.607\pm0.015$ & $0.568\pm0.015$ & $0.510\pm0.027$ & $0.488\pm0.028$ \\
$[0.75,2.0]$ & $0.600\pm0.016$ & $0.560\pm0.014$ & $0.491\pm0.029$ & $0.472\pm0.029$ \\
\hline
\end{tabular}
\parbox{0.94\textwidth}{The normalization is fixed by the initial
state at $|y|=10a$. Errors are one-standard-error log--log fit
uncertainties across four resistivities.}
\end{table*}

\begin{figure*}[!htbp]
\centering
\includegraphics[width=0.95\textwidth]{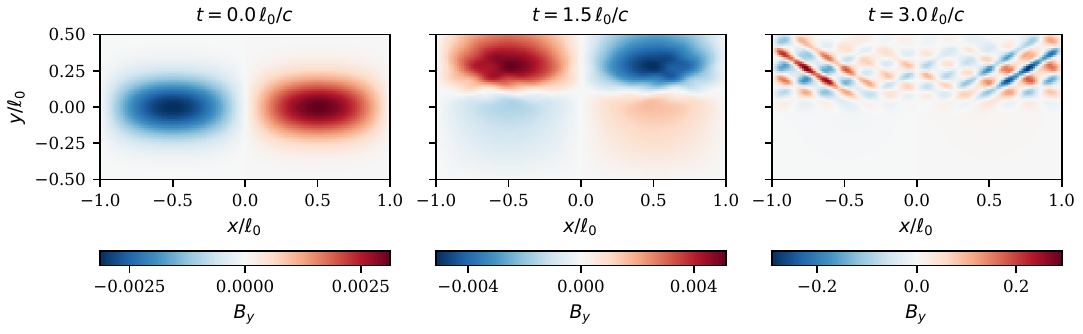}
\caption{Perturbation component $B_y(x,y)$ at $t=0$, $1.5$,
and {$3\,\ell_0/c$}, with independent colorbars per panel. The
reconnecting Harris field $B_x$ appears in
Figure~\ref{fig:harris_ic}. The transverse-field pattern evolves
from the imposed divergence-free perturbation.}
\label{fig:harris_structure}
\end{figure*}

Figure~\ref{fig:harris_structure} shows the $B_y(x,y)$ structure at
three times ($t=0$, $1.5$, $3$), with an independent color scale
in each panel to resolve the evolving transverse field.

\begin{figure}[!htbp]
\centering
\includegraphics[width=0.47\textwidth]{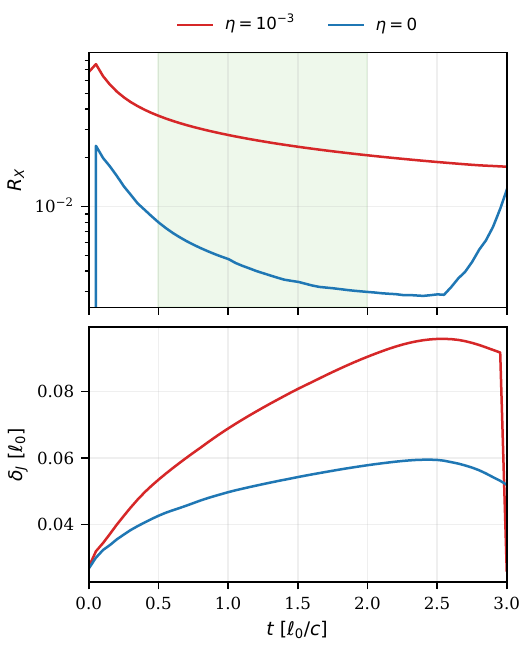}
\caption{{Harris structural run and the same-resolution ideal
reference.} Top: cell-reconstructed local rate with the fixed
initial relativistic Alfv\'en normalization; shading marks
$t=0.5$--$2.0$. Bottom: current-profile thickness $\delta_J$,
defined in the text. Red denotes $\eta=10^{-3}$ and blue denotes
$\eta=0$. The displayed interval is $0\leq t\leq3$ in code
units.}
\label{fig:harris_comparison}
\end{figure}

Figure~\ref{fig:harris_comparison} compares the structural
$\eta=10^{-3}$ case with an ideal reference on the same
$128\times64$ grid. Both rates use the initial relativistic
upstream normalization at $|y|=0.2$, consistent with the matrix.
Over $t=0.5$--$2.0$,
the explicit rate exceeds the ideal-reference rate by the factor
reported in Section~\ref{sec:vandv}.

\begingroup 
For the thickness diagnostic, we interpolate $|J_z|$ onto the
$x=0$ line and locate its two half-maximum crossings by linear
interpolation. We define
\begin{equation}
 \delta_J=\frac{{\rm FWHM}(|J_z|)}{2\operatorname{arcosh}\sqrt2}.
\end{equation}
This equals the scale parameter of a $\operatorname{sech}^2$
current profile. If crossings cannot be found, the
measurement uses $\int|J_z|dy/(2|J_z|_{\max})$, followed by
$B_{\rm up}/|\partial_yB_x|$ if needed.
No layer half-length or steady-state Sweet--Parker rate is inferred
from these curves.
\endgroup 

%-----------------------------------------------------------------------
\bibliographystyle{aasjournalv7}
\bibliography{refs}

\end{document}